\documentclass[aps,pre,twocolumn,superscriptaddress]{revtex4-1}%
\usepackage{epsfig,amssymb,amsmath,amsthm,amsfonts,amsbsy,mathrsfs,bm}
\usepackage{graphicx}
\usepackage{color}
\usepackage{siunitx}
\usepackage{bm}

\usepackage{tikz}
\usepackage[english]{babel}
\usepackage[colorlinks=true]{hyperref}

\begin{document}
\title{In-silico modeling of early-stage biofilm formation}

\author{Pin Nie}
\affiliation{Division of Physics and Applied Physics, School of Physical and Mathematical Sciences, Nanyang Technological University, Singapore 637371, Singapore}

\author{Francisco  Alarcon  Oseguera}
\affiliation{
Departamento de Estructura de la Materia, Fisica Termica y Electronica, Facultad de Ciencias Fisicas, Universidad Complutense de Madrid, 28040 Madrid, Spain
}
\affiliation{
Departamento de Ingeniería Física, División de Ciencias e Ingenierías, Universidad de Guanajuato, Loma del Bosque 103, 37150 León, Mexico
}

\author{Iv\'an L\'opez-Montero}
\affiliation{
Instituto de Investigaci\'on Hospital 12 de Octubre (i+12), 28041 Madrid, Spain
}
\affiliation{
Departamento de Química F\'isica, Universidad Complutense de Madrid, 28040 Madrid, Spain
}

\author{Bel\'en Orgaz}
\affiliation{Departamento de Farmacia Galénica y Tecnología Alimentaria, Universidad Complutense de Madrid,
28040 Madrid, Spain}

\author{Chantal Valeriani}
\affiliation{
Departamento de Estructura de la Materia, Fisica Termica y Electronica, Facultad de Ciencias Fisicas, Universidad Complutense de Madrid, 28040 Madrid, Spain
}

\author{Massimo Pica Ciamarra}
\email{massimo@ntu.edu.sg}
\affiliation{Division of Physics and Applied Physics, School of Physical and Mathematical Sciences, Nanyang Technological University, Singapore 637371, Singapore}
\affiliation{
CNR--SPIN, Dipartimento di Scienze Fisiche,
Universit\`a di Napoli Federico II, I-80126, Napoli, Italy
}
\date{\today}

\begin{abstract}
Several bacteria and bacteria strands form biofilms in different environmental conditions, e.g. pH, temperature, nutrients, etc.
Biofilm growth, therefore, is an extremely robust process.
Because of this, while biofilm growth is a complex process affected by several variables, insights into biofilm formation could be obtained studying simple schematic models.
In this manuscript, we describe a hybrid molecular dynamics and Monte Carlo model for the simulation of the early stage formation of a biofilm, to explicitly demonstrate that it is possible to account for most of the processes expected to be relevant.
The simulations account for the growth and reproduction of the bacteria, for their interaction and motility, for the synthesis of extracellular polymeric substances and Psl trails.
We describe the effect of these processes on the early stage formation of biofilms, in two dimensions, and also discuss preliminary three-dimensional results.
\end{abstract}

\maketitle
Biofilms are self-organized bacteria communities comprising the bacteria and a matrix of extracellular polymeric substances (EPS)~\cite{Vert2012}. 
Biofilms are certainly the most resilient form of life on Earth, as they survive in both hot, salty, acid and alkaline waters, as well as at extremely low temperature. 
Biofilms colonize their host environment, including humans, in which case they are frequently the cause of persistent infections.
Their resilience mainly originates from the EPS matrix,
which might account for up to 90\% of the dry biofilm weight.
Besides allowing for a spatial and social supracellular organization~\cite{Flemming2016}, the matrix provides a physical scaffold that keeps the cells together and protects them from antimicrobial compounds (antibiotics)~\cite{World2014}. 
EPS also play a prominent role in the early stage biofilm formation, by promoting the attachment of bacteria on surfaces~\cite{Berne2018}. 

The social need for research in biofilms is enormous. 
Biofilm grows on the surface of a tooth, causing dental plaque~\cite{Marsh2006}. 
More worryingly, they grow on medical devices~\cite{Francolini2010} such as prosthetic heart valves, orthopaedic devices, skull implants, and might trigger virulent rejection reaction. 
Pseudomonas aeruginosa, for example, can enter the blood circulation~\cite{Micek2005} through open wounds to infect organs of the urinary and respiratory systems. 
In a different context, biofilm cause billions of dollars in damage to metal pipes in the oil and gas industry~\cite{Xu2015,Ashraf2014}. 
Sulfate-reducing bacteria~\cite{Enning2014}, for example, transform molecular hydrogen into hydrogen sulfide which, in turn, produces sulfuric acid that destroys metal surfaces causing catastrophic failures. 
In the water supply system, biofilm can grow in pipes, clogging them due to their biomass~\cite{Mazza2016}. 
It is of enormous interest to develop surfaces to which bacteria are not able to attach. To date, no surface able to reliably inhibit the formation of biofilms is known~\cite{Montanabook1995}.

On the other hand, one might also tame biofilms to benefit from them.
For example, we could exploit biofilms in environmental biotechnology, e.g., in wastewater treatment~\cite{Lazarova1995}, or for in situ immobilization of heavy metals in soil~\cite{Flemming1996}. 
Biofilms naturally grow by consuming organic materials in the fluid. 
Microorganisms (typically bacteria and fungi) can be used for microbial leaching, e.g., to metals from ores. Copper, uranium, and gold are examples of metals commercially recovered by microorganisms~\cite{Mazza2016}.

The life cycle of a biofilm is traditionally described as consisting of five phases: reversible attachment, irreversible attachment, growth, maturation and dispersion. 
The first three phases identify the early-stage biofilm formation.
Understanding this phase is of particular interest, as it might allow for the design of mechanisms able to prevent the formation of a biofilm.
There is mounting evidence that, in this phase, mechanical forces play a crucial role in this stage~\cite{Allen2019}, affecting the growth dynamics as bacteria diffuse on the surface to be colonized, interacting among themselves and with a chemical environment affected by their secretions.
These include EPS, and in particular, Psl exopolysaccharide, which promotes surface attachment.

The observation that biofilms are formed by different bacteria and bacteria strands, under highly variable external conditions, suggests that schematic models could provide critical insights into biofilm formation.
Indeed, several models have been introduced in the literature~\cite{Rudge2012,Winkle2017,Mattei2018}, e.g. to investigate biofilm jamming~\cite{Delarue2016}, nematic ordering~\cite{DellArciprete2018,Acemel2018}, role of psl trails~\cite{Zhao2013}, nutrient concentration~\cite{Rana2017}, phase separation~\cite{Ghosh2015}, front propagation~\cite{Farrell2017}.

In this manuscript, we introduce a flexible computational model for the investigation of the early-stage biofilm formation.
As in previous models, we describe a biofilm as a collection of growing and self-replicating rod-shaped particles. 
We do, however, also consider the role of Psl trails reproducing previous experimental results~\cite{Zhao2013}, and model for the first time the growth of an EPS matrix, 
The article is structure as follows.
In Sec.~\ref{sec:numerical} we introduce the numerical model, detailing all of the features we consider as well as those we decided to neglect.
We then examine the behavior of the model, investigating different scenarios in increasing order of complexity: Growth of non-motile cells, Sec~\ref{sec:one_growth}; competition between growth rate and motility, Sec.~\ref{sec:mot_growth}; multi-species biofilm, Sec.~\ref{sec:twospec}; role of Psl trails~\ref{sec:psl}; formation of the EPS matrix, Sec~\ref{sec:eps}. 
We conclude discussing the transition from two- to three-dimensional colonies~\ref{sec:3d}, and future research directions.

\section{Numerical model~\label{sec:numerical}}
Modelling the biofilm early-stage formation is a challenging task, as one need to accounts for several biological and out-of-equilibrium processes.
The microscopic model also needs to be supplemented by several parameters, e.g. to describe motility, reproduction, eps production, etc. 
We describe in the following the main features of the computational model we have implemented. 
While the model is general, we have calibrated the values of its many parameters by referring to previous experimental investigation of the pathogen {\it Pseudomonas aeruginosa}, whenever possible.

\begin{figure}[!!t]
\centering
\includegraphics[width=0.5\textwidth]{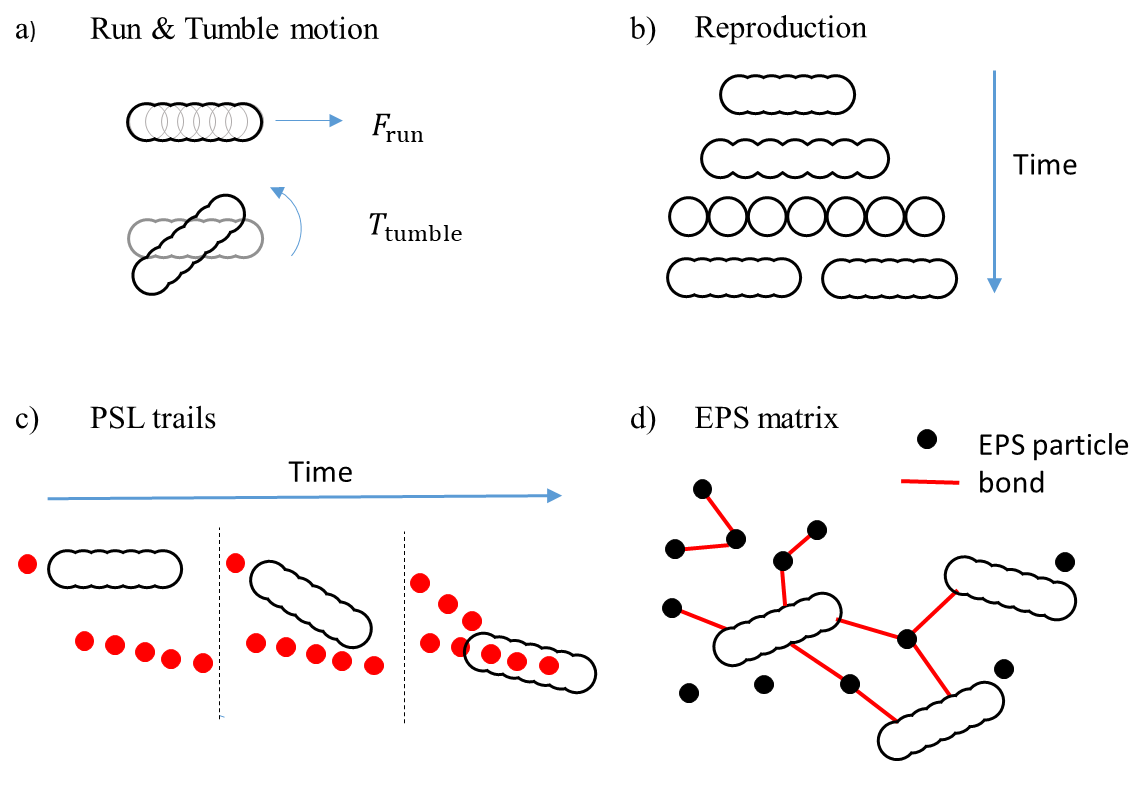}
\caption{Schematic illustration of the of the considered model. 
a) Bacteria are modeled as a collection of particles. Isolated bacteria undergo a run and tumble motion, we realize adding a  propelling force and a torque, in a viscous background. 
b) Consecutive particle making a bacterium interact via a harmonic spring of rest length $l_0$. 
We model bacterial grow making $l_0$ time dependent. 
A bacterium reproduces when its size doubles. 
c) Bacteria may deposit a psl trail (red dots) as they move on the surface. These immobile psl particles attract the particles making up a bacterium, effectively exerting a net force and torque.
Because of this, moving bacteria preferentially follow existing psl trails.
d) Bacteria may produce eps, we model as small particles. Permanent bonds are formed between the EPS particles, and between the EPS particles and those making up the bacteria. This polymerization process leads to the formation of a EPS matrix.
\label{fig:schematic}}
\end{figure}

We describe in the following the implementation of different features of the model, in order of complexity, which are schematically illustrated in Fig.~\ref{fig:schematic}.

\subsection{Isolated non-reproducing bacterium}
We model a bacterium as a spherocylinder, which we construct by lumping together $7$ point particles. 
Point particles of different bacteria interact via a Weeks-Chandler-Anderson (WCA) potential.
This is a Lennard-Jones potential with energy scale $\epsilon$ and diameter $\sigma$, we cut at its minimum $d_{\rm b,b}=2^{1/6}\sigma$.
This distance fixes the transverse width of the bacteria that, in our units, is $w = d_{\rm b,b} = 0.6\mu$m. 
Consecutive particles of a bacterium interact via a Harmonic spring with stiffness $k_{\rm b} = 250\epsilon/w^2$ and initial rest length $l_0$, we fix so that the bacterium aspect ratio is $[(n-1)l_0+w]/w = 3$.
These value for the size of a bacterium mimic that of {\it Pseudomonas aeruginosa}.
Bending rigidity is provided introducing Harmonic angular interactions, with rest angle $\pi$ and stiffness $k_{\rm a} = 20\epsilon$, between any three consecutive particles. 
The value of the stiffness coefficient is high enough for the bending deformation of the bacteria to be negligible, for the range of parameters we will consider.

We assume the bacteria to follow an overdamped dynamics, which we realize by applying to each particle making up a bacterium a viscous force $-\gamma v$ proportional to its velocity. 
Here $\gamma$ is a viscous friction coefficient.
We further assume the bacteria to perform a run and tumble motion. During a `run' period, whose duration is a random number drawn from an exponential distribution with time constant $t_{\rm run} = 3$ min, we apply to the particles making a bacterium a force $F = v_{\rm run}/\gamma$, where $v_{\rm run} = 0.12 \mu m/s$ is the velocity of the particles in the running state. 
During a `tumble' period, whose duration is a random number drawn from an exponential distribution with time constant $t_{\rm tumble} = 0.5$ min, we apply to the bacterium a torque $T$, which fixes a rotational velocity. 
The equations of motion are solved with a Verlet algorithm with timestep $5\cdot 10^{-3}s$.
The dynamical properties of a bacteria depend on the species, mutant, as well as on the experimental condition.
The values described above reasonable reproduce the time dependence of the mean square displacement curves of Ref.~\cite{Conrad2011}, conducted in the early stage of formation of P. aeruginosa biofilms. In particular, the diffusion coefficient results $D \simeq 0.7 \mu^2/s$.

\subsection{Growth and reproduction} 
We model the growth of bacterium by making time-dependent the rest lengths of the springs connecting the beads making-up bacterium. 
Precisely, the rest lengths grow linearly in $\min(t-t_{\rm b},1.2t_r)$, where $t$ is the actual time and $t_{\rm b}$ the time of birth of the bacterium, with a grow rate set such that an isolated bacterium double its length in $t_r$, where for each bacterium $t_r$ is taken from an exponential distribution with mean $\langle t_r \rangle = 1$h. 
The maximum value of the rest length has a cutoff to avoid the unbounded growth of the pressure of a bacterium not able to grow, e.g. as in a dense environment.
A bacterium reproduces when its length equals twice the original one.
We implement the reproduction by replacing a bacterium with two daughter cells, which occupy the same volume as the original one. The polarity of the daughter cells is that of their father.

\subsection{Psl exopolysaccharide trails} 
When moving on a surface, bacteria may secrete Psl exopolysaccharide.
Psl promotes attachment, effectively acting as a glue~\cite{Zhao2013}.
Describing this process requires keeping track of the spatial location visited by the moving bacteria.
From a computational viewpoint, we do that by superimposing to the computational domain a square grid, with grid size $l \simeq w/20$, where $w$ is the width of the bacteria.
As the bacteria move on the surface, we record how many times each cell is visited.
Specifically, considering our coarse-grained description of the bacteria as a collection of particles, we focus on the position of the central one.
We indicate with $n_v({\bf r},t)$ the number of times the grid cell in ${\bf r}$ has been visited; this number originates from the superimposition of the trails left by all bacteria.
We assume $n_v({\bf r},t)$ to be proportional to the amount of Psl deposited by the bacteria in ${\bf r}$. 

To model the interaction between the bacteria and the trail pattern, we add to the energy of our model the following term:
\begin{equation}
    V_{\rm trail}(t) = \sum_b \sum_{r_i \in b} \sum_r  n_v({\bf r},t) v_{\rm Gauss}({\bf r}- {\bf r_i}),
\end{equation}
where the first sum runs over all bacteria, the second one over the particles of a bacterium, and the third one over the cells of the grid we use to record the trail pattern. 
The interaction between each cell element and each particle of our bacteria is given by an attractive potential, whose amplitude is proportional to the number of times the grid element has been visited. 
We model this attractive potential with an attractive Gaussian potential  $v_{\rm Gauss}$, with a width equal to half of the bacterial width.
Notice that the trail interaction acting on each bacterium exerts a torque, whose net effect is that of aligning the bacteria to the trail.

In this model, the interaction potential is characterised by a typical energy scale, $\epsilon$. We do not find literature data discussing the strength of this interaction. Also, the rate of which bacteria deposit Psl has not been discussed in the literature. 
Nevertheless, we understand that if bacteria deposit Psl too frequently, and if the attraction is too strong, then the bacteria will quickly bind to the deposited Psl, and will stop diffusing~\cite{Tsori2004,Sengupta2009}. 
This self-trapping appears to be unrealistic.
On the order side, if the deposition rate is too small, then the bacteria deposit Psl in uncorrelated locations, not on a trail. This scenario also appears unrealistic.
We have, therefore, arbitrarily chosen simulation parameters for which the concept of a trail is well defined.

\subsection{Extracellular Polymeric Substances} 
EPS production is essential to the growth of biofilm in vivo, as it bridges bacteria cell together and to the hosting surface~\cite{Xiao2009}. 
In the early stage formation, EPS production appears to cooperate with bacterial motility, e.g. twitching motility~\cite{Conrad2011}, as bacteria need to be close in space to agglomerate.
Indeed, motility suppression may hinder the formation of microcolonies and biofilms~\cite{Recht2000}, at least if the bacteria do not explore their environment via other physical processes, e.g. diffusion or drift in a flow. 

The theoretical and numerical description of the role of EPS is arduous and limited. 
Here, we develop a numerical model for EPS along the line of the only literature work explicitly modelling EPS particles~\cite{Ghosh2015} we are aware of, but also introducing substantial advancements.
Considering EPS as polymer coils, Ref.~\cite{Ghosh2015} has modelled EPS as point particles interacting via a purely repulsive potential.
These particles have been considered as passive and not able to form bonds to give rise to an EPS matrix.
In this condition, EPS and bacteria have been found to phase separate, a result rationalized invoking a depletion-like interaction~\cite{Ghosh2015}.
Regardless, the features of the observed phase separation depend on the rate at which EPS particles are produced. 
More recent results have also highlighted the interplay between motility and depletion-like interactions~\cite{Porter2019}.

The main novelty of our approach is in the introduction of a polymerization dynamics, allowing EPS particles to bond among themselves and with the bacteria, to create an EPS matrix.
Specifically, we describe EPS particles and their dynamics as follows:
\begin{enumerate}
    \item Extracellular polymeric substances (EPSs) are represented as small spheres, whose size is half of the width of the bacteria, $\sigma_{\rm eps} = D/2$.
    \item  EPS particles interact among them with a purely repulsive WCA potential, with energy scale $\epsilon$, as the particles of different bacteria.
    \item EPS particles are inserted by the bacteria in their surrounding, at a rate $\tau_{\rm eps}^{-1}$. An EPS particle is inserted only if it does not interact with any other particle or bacteria.
    This ensures numerical stability. Hence, EPS production is suppressed in crowded conditions.
    \item Every $\Delta t$, where $\Delta t$ is a random variable taken form an exponential distribution with average value $\Delta_t^*$, we look for all possible pair of interacting EPS particles.
    If two EPS particles are interacting, we add an harmonic bond $v(r) =  10^2\epsilon (r-\sigma_{\rm eps})^2$ between them, provided that they are not already bonded, with a probability $p_b$.
    \item Similarly, every $\Delta t$ we add a bond between an EPS particle and a bacteria particle in interaction, provided that they are not already bonded, with probability $p_b$. 
    In this case, the bond energy is $v(r) = 10^2 \epsilon \left[r- \left( \frac{\sigma_{\rm eps}+D}{2}\right) \right]^2$.
\end{enumerate}
The steps 1-3 above essentially reproduce the model of Ref.~\cite{Ghosh2015}.
On the other hand, steps 4-5 describe the dynamics of a polymerization process. 
The ratio between the mass $m_{\rm eps}$ of an EPS particle and the mass $M$ of a bacterium is $m/M \ll 1$.
EPS particles motion follow a Langevin dynamics, with parameters fixed so that a particle has thermal velocity $\sqrt{2k_BT/m_{\rm eps}} = 0.18\mu$m/s, and a diffusion coefficient roughly 100 time smaller than that of bacteria in dilute conditions. 
This means that the bacteria de-facto move in a bath of almost immobile EPS particles.

The EPS model has two parameters, $\Delta_t^*$ and $p_b$, and the rate at which bonds are formed between possible pair of particles is $p_b \Delta_t^*$. 
It isn't easy to estimate these parameters from the experiments. 
Besides, we notice that the EPS production rate depends on the growing condition.
Here, we decided to fix $\Delta_t^* = 1$min $=\tau_r/60$, and have investigated the dependence of the growing dynamics on the bond probability $p_b$.
We consider the bond between bacterial and EPS particles to be permanent.

\subsection{What is not in the model}
This model takes into account all of the processes that appear to be relevant, such as motility, reproduction, production of Psl trail, EPS matrix, etc.

Some features, we believe to be less relevant, are for now neglected.
For instance, we neglect hydrodynamic interactions, which after the initial docking of the bacteria should be minor, due to the small Reynolds number.
Indeed, bacteria swim in bulk with velocity $\simeq 30 \mu m/s$, and on surface with velocity $\simeq 1 \mu /s$. 
The Reynolds number is ${\Re} = \frac{\rho_f  v L}{\nu}$, where $\rho_f$ is the density of the fluid, $\nu$ is its viscosity ($\nu = 10^{-3}Pa s$ for water), $v$ is the relative velocity of the particle with respect to the fluid, $L$ is the typical length of a bacterium (around $1 \mu m$). 
Thus, for bacteria swimming in bulk, the Reynolds number is $\sim 3 \times 10^{-5}$, and for Bacteria on the surface, the Reynolds number is $\sim 10^{-6}$. 
Bacterial motion is thus in a low Reynolds number regime where viscous forces dominate over inertial ones. 

Furthermore, we do not consider the diffusion of nutrients and hence the possibility that the growth rate and the motility properties might spatially vary. 
In the early-stage formation in which the biofilm is essentially two-dimensional, we do not expect diffusion of nutrients to be sensibly affected by the forming biofilm. 
Indeed, experimental results suggest that the growth rate in the interior and the periphery of a biofilm are comparable~\cite{Zachreson2017}.

\section{Growth in the absence of motility \label{sec:one_growth}}
\begin{figure}[!!t]
\centering
\includegraphics[width=0.5\textwidth]{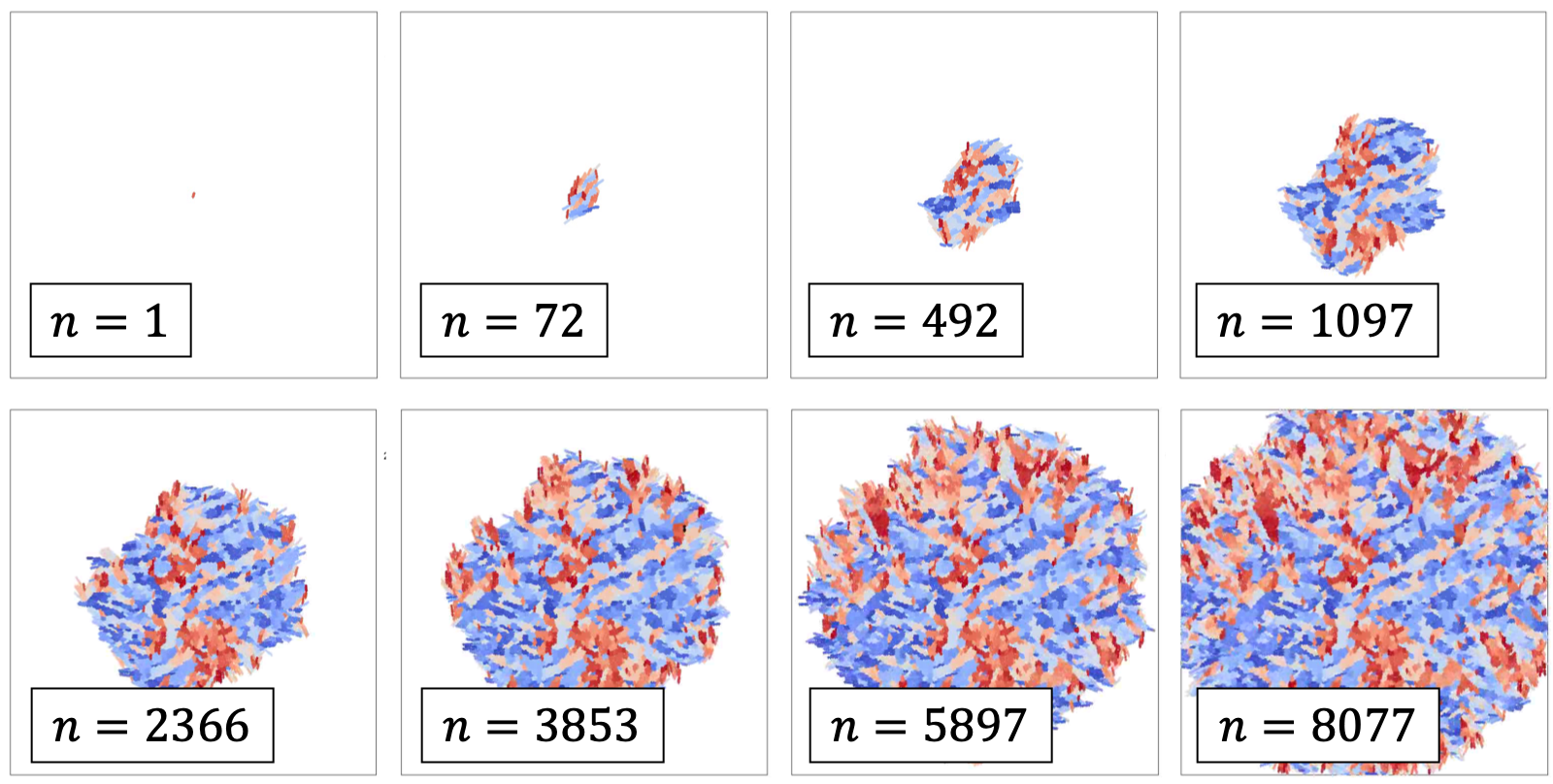}
\caption{Growth of a colony of non-motile bacteria, imaged every $4$h.
The colour code reflects the angle between the bacteria and a fixed spatial direction. Hence, patches with the same colour correspond to regions with the same nematic director. 
See \href{https://youtu.be/_brD7-XXBLU}{here} for the corresponding animation.
\label{fig:nematic}}
\end{figure}
We begin illustrating our model at work with the simplest possible example. 
The growth of a colony of non-motile bacteria, in the absence of Psl and EPS. 
In this scenario, we do expect the number of bacteria to grow exponentially with time. 
Saturation occurs at large times due to finite-size effects. 
This jamming transition induced by reproduction has been considered before~\cite{Delarue2016}.

We illustrate the expanding colony in Fig.~\ref{fig:nematic}, where a fixed time interval separate consecutive snapshots.
The number of bacteria $n$ present at each time is specified in each panel. The direct visualization of the colony suggests that the bacteria tend to align with each other. 
Nematic ordering is indeed commonly observed in experiments~\cite{Volfson2008,DellArciprete2018,Yaman2019}; the order is short-ranged due to the emergence of buckling instabilities~\cite{Boyer2011}.

To investigate this issue, we colour code each bacterium according to the angle its director forms with a given axis (modulus $\pi$, given that in the absence of motility the bacteria are not polar). 
As the colony grows, we see the emergence of domains with the same colour corresponding to regions of local nematic alignment.

\section{Motility vs. growth rate\label{sec:mot_growth}}
\begin{figure}[!!t]
\centering
\includegraphics[width=0.5\textwidth]{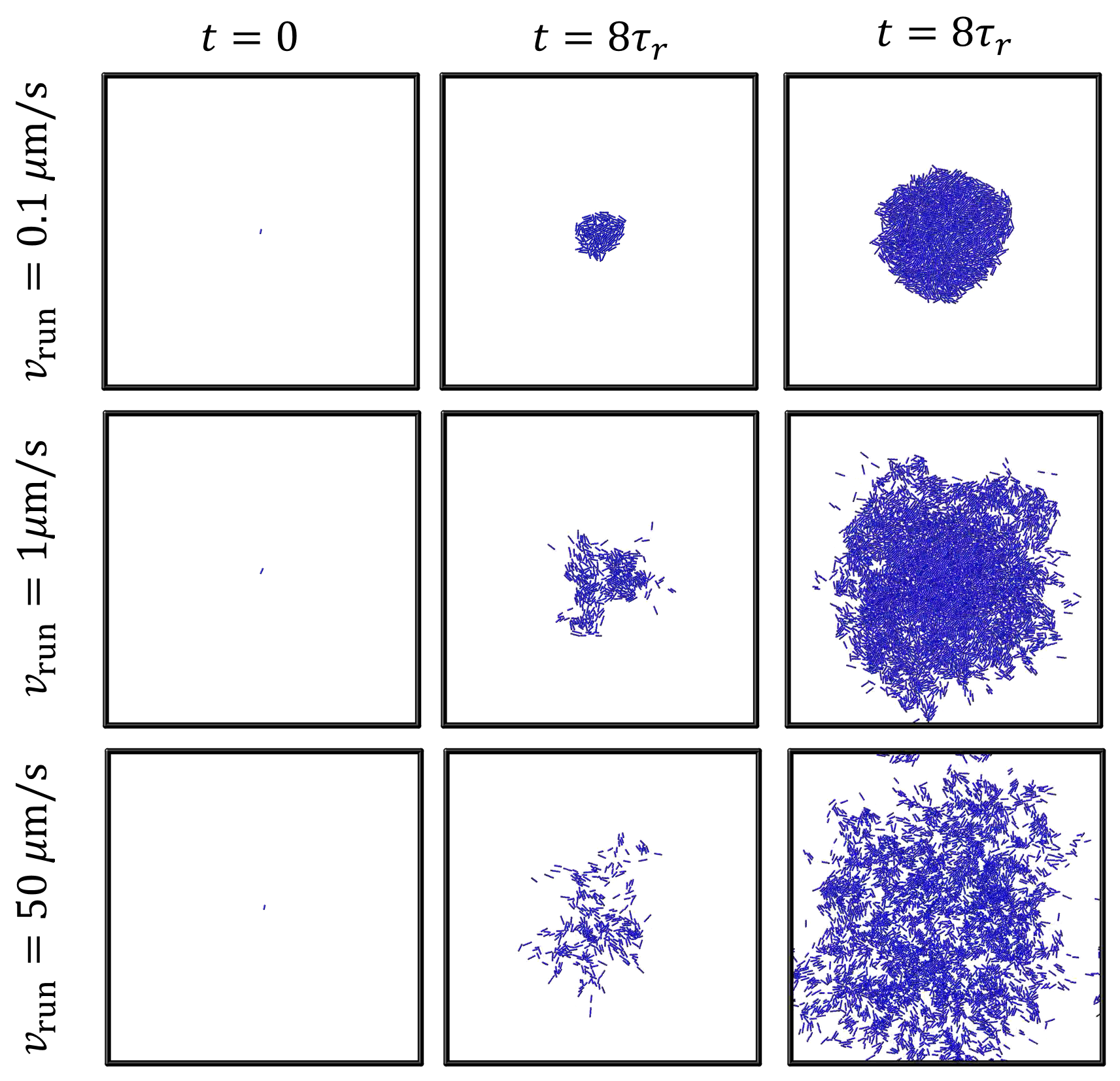}
\caption{Growth of microcolonies of bacteria having different typical velocity $v_{\rm run}$ and fixed average reproduction time, $\tau_{r} = 1$h. See these links for the corresponding animations: \href{https://youtu.be/IBO6xreaKe8}{slow}, \href{https://youtu.be/eZdjqvk7BSY}{medium},
\href{https://youtu.be/b1epy1512As}{fast}.
\label{fig:DiffMobility}}
\end{figure}
The motility properties of bacteria are highly variable.
Different species have different motility properties.
For each species, motility depends on the mutant, e.g. depending on the presence of type-4 pili or of the flagella.
Besides, motility depends on the external environment, e.g. on the presence of nutrients.
Because of this variability, it is interesting to consider the dependence of the early-stage formation on the motility properties, in our numerical model.

Here we consider that, once a bacterium adheres to the surface and seeds a microcolony, the subsequent evolution depends on the competition of two physical processes,  reproduction and motility.
To clarify the origin of this competition, we start by considering the time dependence of the radius of a microcolony, assuming the bacteria to have no motility.
In this condition, a colony expands as bacteria duplicate and push against each other.

To model this situation, we assume the colony to have a constant number density $\rho$, number of bacteria per unit area, so that then the number $n$ of bacteria in a colony of radius $R$ is $n(R) = \rho 4\pi R^2$. 
How does $R$ evolves with time?
To predict $R(t)$, we assume the bacteria to reproduce with a constant rate $\tau_r^{-1}$, so that $\frac{dn}{dt} = \frac{n}{\tau_r}$.
From this assumption, we get
\begin{equation}
\frac{n}{\tau_r} = \frac{dn}{dt} = 8 \pi \rho R \frac{dR}{dt}.
\end{equation}
Hence, the radial expansion velocity of the colony is
\begin{equation}
v_R = \frac{dR}{dt} = \frac{R}{2\tau_r}.
\end{equation}
Interestingly, this model predicts that the expansion velocity grows linearly with the cluster size.
One might expect this to occur in the early stage development of a microcolony. 
At a later time, the bacteria deep inside the colony stop reproducing because of the limited nutrient diffusing to the core or because of the high mechanical pressure.

If the bacteria are motile, then another typical velocity scale enters into the problem: the characteristic bacteria velocity $v_{\rm run}$.
It turns out that $v_R$ and $v_{\rm run}$ compete.
Precisely, when $v_{\rm run} \gg v_R$, bacteria swim away from each other before they reproduce. 
Conversely, they reproduce when still close.
Since $v_R$ grows with the bacteria colony, there is a characteristic colony radius $R \simeq 2 v_{\rm run} {\tau_r}$ above which the radial velocity profile due to the reproduction overcomes the swimming velocity of the bacteria. When this occurs, the colony starts becoming compact.

As an example, we illustrate in Fig.~\ref{fig:DiffMobility} the developing of three different microcolonies, which only differ in the magnitude of the typical velocity of bacteria.
At small velocities, the microcolony is nearly compact at all times.
At large velocities, bacteria spread on the surface at short times, as apparent in the configuration reached at $8\tau_r$ in the case of intermediate velocities, but then become part of a dense microcolony. 
At even larger velocities, compact shape is attained at a longer time, possibly not yet achieved in our simulation with $v_{\rm run} = 50$.

It is interesting to notice that, in this picture, a compact colony emerges in this picture when the reproduction rate dominates over the motility of the particles. 
In this respect, while microcolony formation visually resemble the activity drive phase separation of active system of spherical
~\cite{Redner2013,Wysocki2014,Fily2012,Buttinoni2013,Palacci2013,Theurkauff2012,Ginot2018,Nie2020,Nie2020a} or dumbbells particles~\cite{Suma2014,Petrelli2018}, the underlying physical driving force is different.

It is, however, arduous to understand the experimental relevance of these findings. 
Indeed, one might expect that before a compact shape is attained, the colony stops expanding in two dimensions, and start growing in the vertical one. We discuss such a transition in Sec.~\ref{sec:3d}.
Besides, in the picture we are considering, there are no bacteria in the planktonic state joining the colony, and no bacteria move from the colony to the planktonic state. 
We do not consider these processes in our numerical model, despite it would be trivial to include them, as the rates of attachment and detachment have not yet been thoroughly experimentally characterized.

\section{Coexistence of different species\label{sec:twospec}}
\begin{figure}[!!t]
\centering
\includegraphics[width=0.5\textwidth]{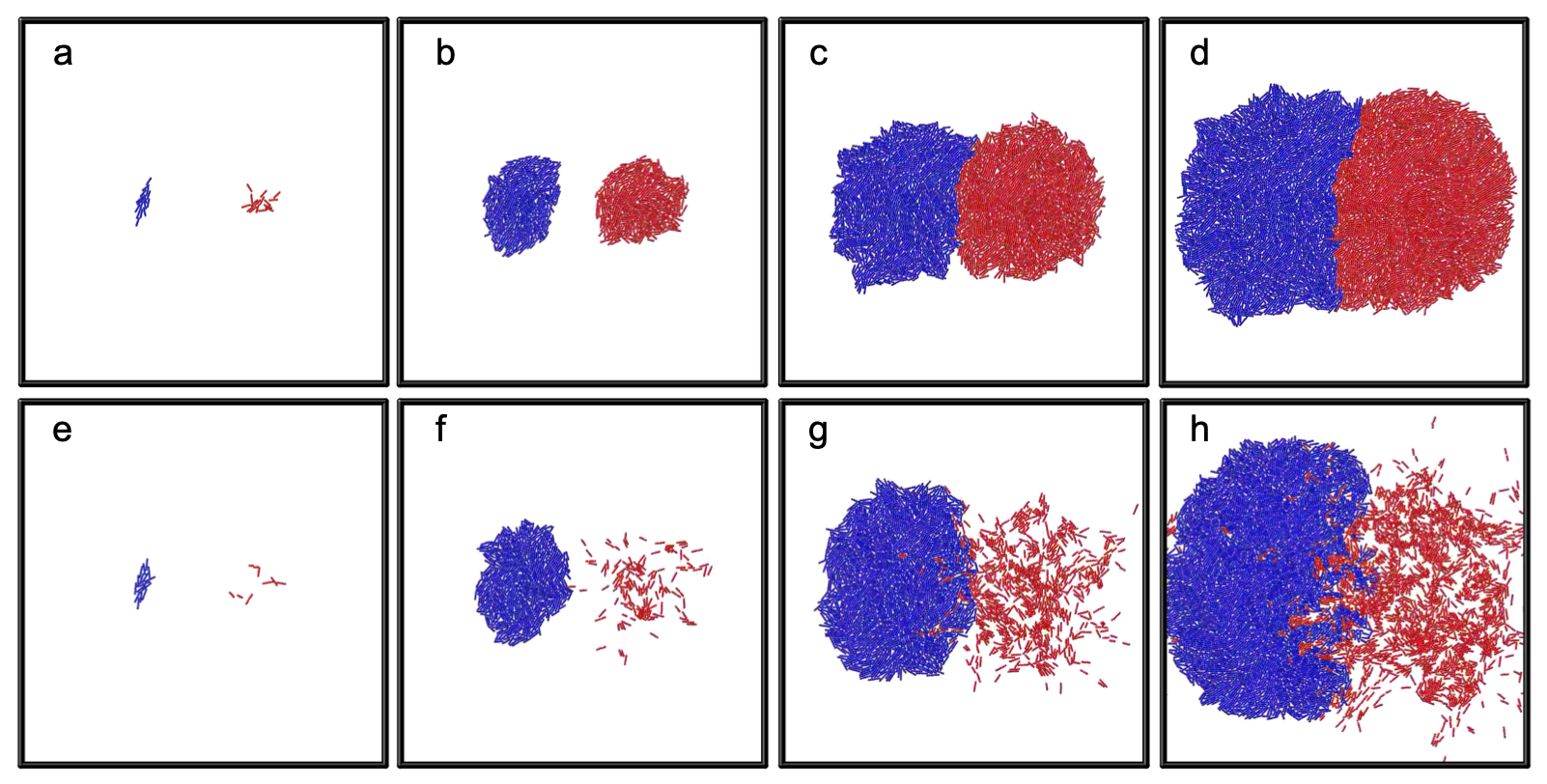}
\caption{Early stage formation of a two-species biofilm. Blu bacteria (left in the figures) are non-motile, while red bacteria are motile (right in the figures). The motile bacteria are faster in the bottom row. See \href{https://youtu.be/754bMRzUSeE}{here} for an animation.
\label{fig:2s2v}
}
\end{figure}
Biofilms are often multispecies~\cite{Roder2016}.
Our computational model allows considering the coexistence of bacteria with different properties.
Here, as an example, we consider that of bacteria with different motility properties.

Fig.~\ref{fig:2s2v} illustrates the growth of a colony of immotile bacteria (blue, on the left), and a colony of motile ones (red, on the right).
On the top row, we consider the case in which the colony of motile bacteria becomes compact before the two colonies start interacting.
Hence, when the two microcolonies enter in contact, both of them are compact. 
As a consequence, a sharp interface between the two colonies develops. 
Notice that this interface is not straight, but slightly curved. 
This curvature reflects the anisotropy of the microcolony of non-motile bacteria, which is ellipsoidal at short times.

In the bottom row of Fig.~\ref{fig:2s2v} illustrates a case in which the motile bacteria are fast so that when the two colonies start interacting, their microcolony is not compact.
In this case, the interface between the two colonies is rough.
A close look suggests that the interface might have a wavy appearance reminiscent of the viscous fingering Saffman–Taylor instability which develops when fluids with different viscosity pushed against each other.
In this respect, we notice that such instability has been reported at the interface of cell populations growing at different rates~\cite{Mather2010}, and in a variety of other contexts~\cite{Ciamarra2005a,PicaCiamarra2005a}.

\section{The role of psl\label{sec:psl}}
\begin{figure}[!!t]
\centering
\includegraphics[width=0.5\textwidth]{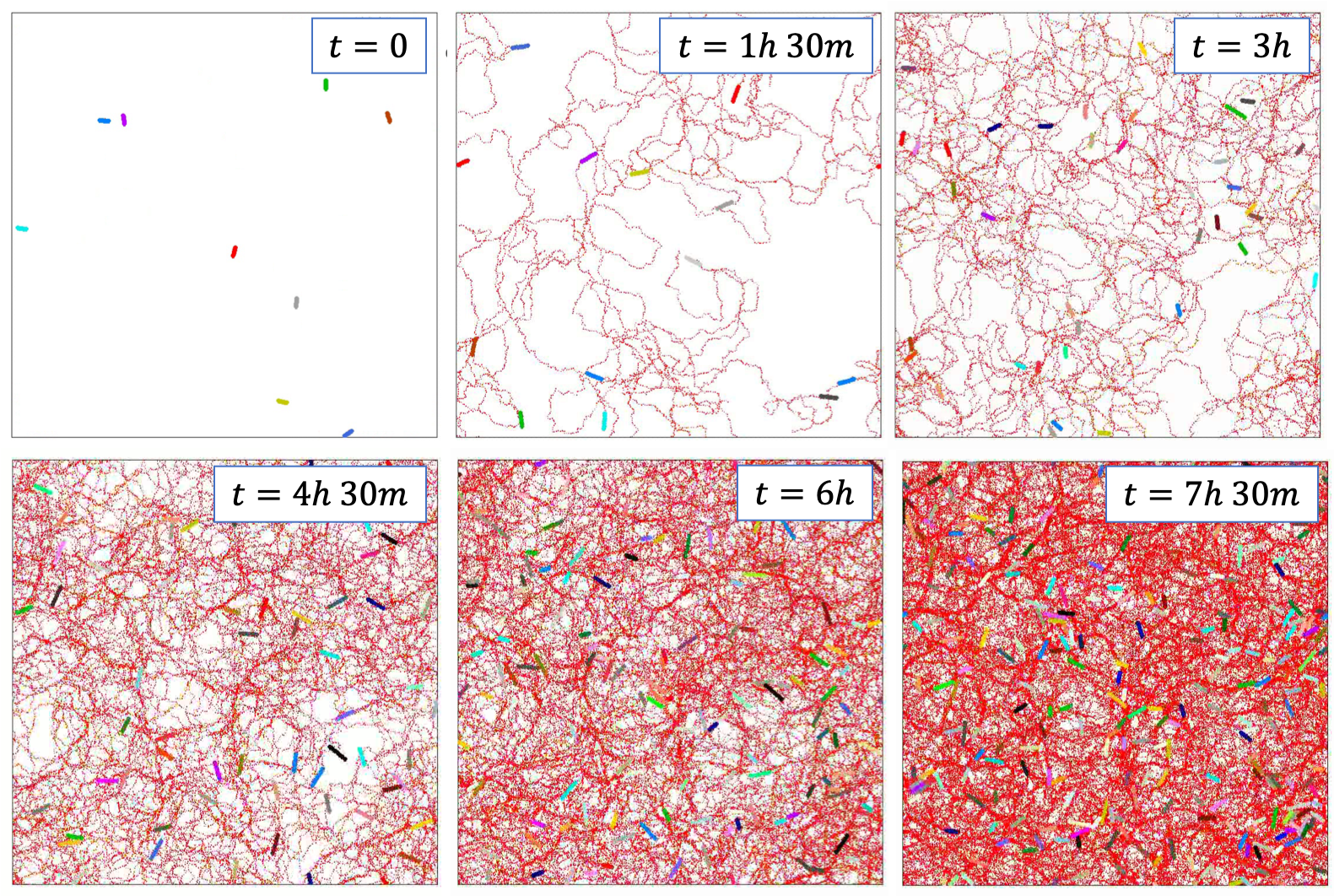}
\caption{Early stage biofilm formation in the presence of Psl production. The red lines are the trails left by the bacteria as they explore the surface. 
Bacteria interact through an attractive force with the trails. The attraction to a particular location is space is proportional to the number of times this location has been visited by the bacteria. See \href{https://youtu.be/lAYZvULn2tk}{here} for the corresponding animation.}
\label{fig:psl_sim}
\end{figure}

While exploring a surface, bacteria may leave a Psl trail, to which other bacteria are subsequently attracted.
Psl trails thus resemble pheromones trails left by ants.
The statistical features of the motion of particles attracted by substances they secrete, generally known as reinforced random walks, have been extensively investigated in the literature~\cite{Allen2019}.
For the case of a single bacterium attracted to its own secreted substance, for instance, Tsori and de Gennes~\cite{Tsori2004} suggested the presence of self-trapping in one and two spatial dimensions, not in three. More recent numerical simulations indicate that there is no self-trapping, but rather a prolonged sub-diffusive transient~\cite{Sengupta2009}. 
Here, we consider the growth of a microcolony, seeded by a single bacterium, in the presence of Psl production.

In Fig.~\ref{fig:psl_sim}, we illustrate a representative time evolution of a bacterial colony.
Besides drawing the bacteria, we illustrate the corresponding trails, which are clearly visible at short times, before trails of different bacteria overlap. 
Qualitatively, these results are analogous to that experimentally reported in Ref.~\cite{Zhao2013}.

\begin{figure}[!!t]
\centering
\includegraphics[width=0.48\textwidth]{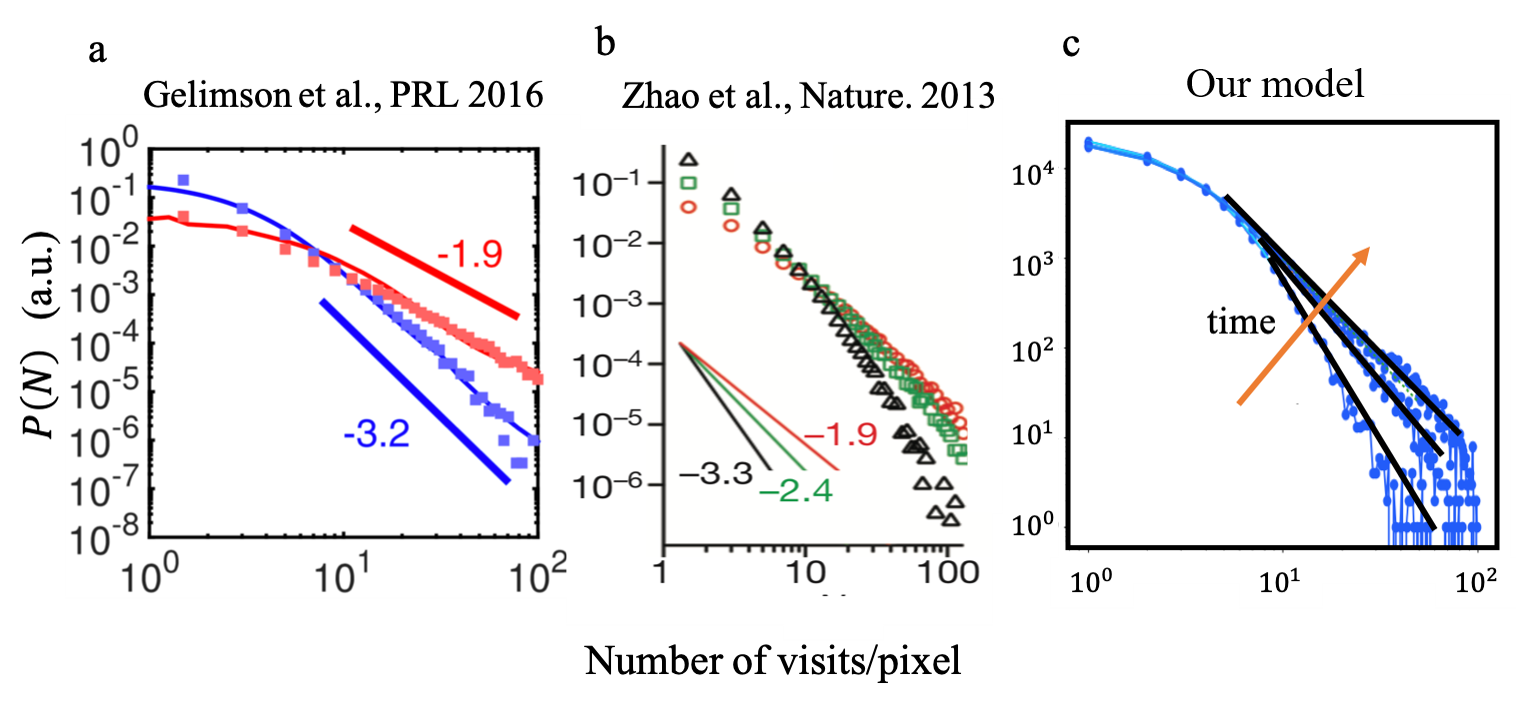}
\caption{Experimental and numerical results for the time evolution of the probability distribution of the number of times a point (pixel) has been visited by a bacterium. Panels a and b report experimental results from Ref.~\cite{Gelimson2016} (with permission) and Ref.~\cite{Zhao2013} (with permission), respectively.
Panel c illustrates the results of our numerical model.}
\label{fig:psl_com}
\end{figure}

To be more quantitative, we have determined the time evolution of the probability distribution of the number of times a particular space location has been visited.
Here, by location, we intend grid elements of side length equal to 1/20th of the bacterial width.
This visit frequency distribution quantity favourably compares to experimental results.
Fig.~\ref{fig:psl_com}a,b presents experimental results for this probability distribution~\cite{Zhao2013,Gelimson2016}. 
The probability distribution decays as a power law, with a large exponent that decreases as times evolve.
In panel c of the same figure, we present our numerical results for the same quantity.
The numerical model well reproduces the experimental results, both as concern the presence of a power-law decay in the probability distribution, as well as the value of the decay exponent and its time dependence.

\section{EPS matrix \label{sec:eps}}
Since EPS come into the focus of the research community only recently, the current knowledge of its role in early-stage biofilm development pales when compared to the extensive understanding of biofilm formation in the absence of EPS production, in particular for non-motile bacteria.
The role of EPS has not been considered in earlier literature\cite{Wingender1999}, as ``traditionally, microbiologists used to study and to subculture individual bacterial strains in pure cultures using artificial growth media. Under these in vitro conditions, bacterial isolates did not express EPS-containing structures or even lost their ability to produce EPS''. 
However, it is nowadays clear that EPS is of fundamental importance, as it allows for a spatial and social supracellular organization~\cite{Flemming2016}, while providing a physical scaffold that keeps the cells together and protect them from antimicrobial compounds and heavy metals~\cite{Nadell2015}, and can also retain water~\cite{Wingender1999}. 
EPS also appears to play a prominent role in the early stage biofilm formation, by promoting the attachment of bacteria on surfaces~\cite{Berne2018}. 

\begin{figure}[!h]
\includegraphics[width=0.5\textwidth]{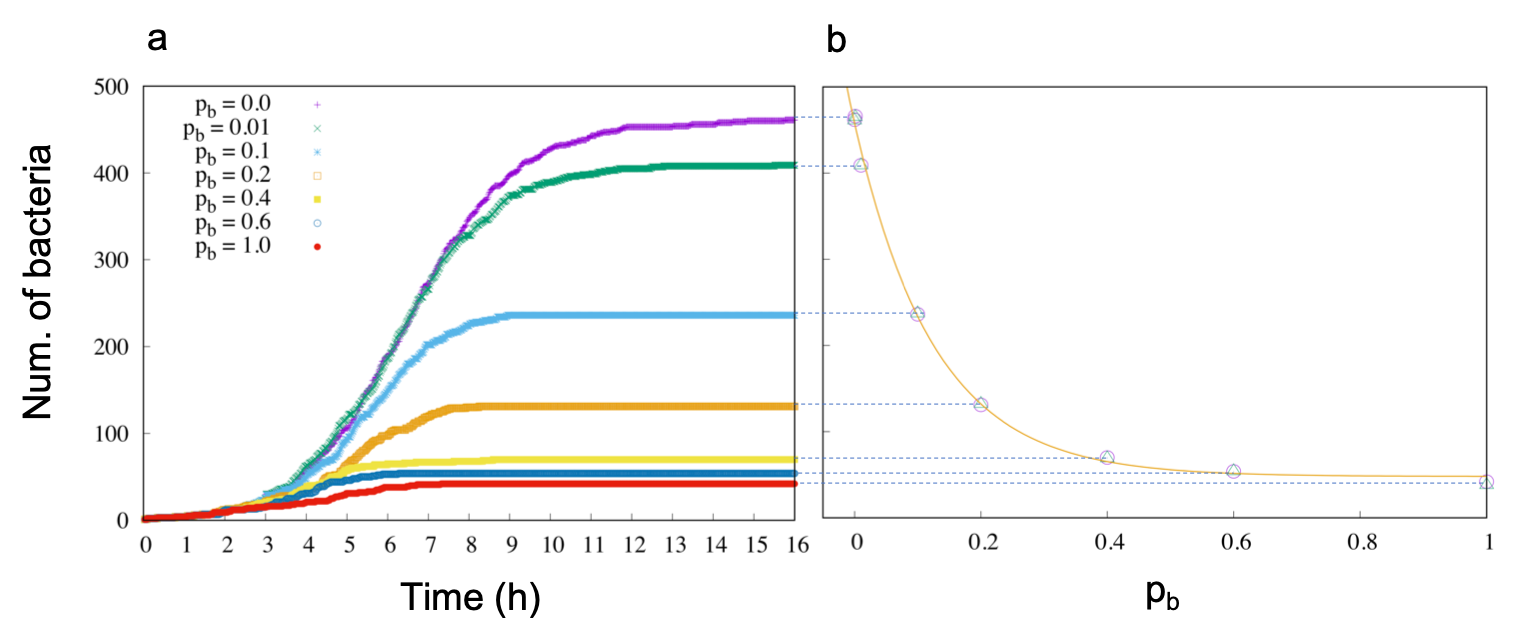}
\caption{Effect of the bonding probability on the number of bacteria. 
Panel (a) illustrates the time dependence of the number of bacteria on the surface. 
Different curves refer to different values of the bonding probability, $p_b$.
Panel (b) shows the dependence of the asymptotic steady state number of bacteria on the bonding probability $p_b$. The fitting line is an exponential one, $n_\infty + (n_0-n_\infty) e^{-p_b/p_b^*}$.
\label{fig:growth_pb}
}
\end{figure}

In our numerical model, two control parameters affect the role of EPS. 
First, there is the rate at which individual bacteria secrete EPS particles in their surrounding, provided that these new particles do not interact with other EPS particles or bacteria. 
We keep this rate to 1/60th or the reproduction rate.
Secondly, there is the probability $p_b$ that two EPS particles, or an EPS and a bacterium, for a bond if close enough. 

Here, we investigate the dynamics and the steady-state as a function of the bonding probability $p_b$. Fig.~\ref{fig:growth_pb}a illustrates the time dependence of the number of bacteria, for different values of $p_b$. 
At short times, $t < 2h$, the production of EPS does not quantitatively affect the dynamics, as different curves collapse on each other.
At larger times, the population grows exponentially but then saturates.
This saturation is not a finite-size effect. 
This is a critical result, as it clarifies that in the presence of EPS a microcolony stops spreading, in two dimensions. 
Indeed, we do expect a transition towards a three-dimensional condition.
Fig.~\ref{fig:growth_pb}b shows that the asymptotic number of bacteria decreases exponentially with the bonding probability. 
If $p_b$ is very high, growth stops with just a few bacteria on the surface. This finding is reminiscent of early speculations for isolated non-reproducing particles~\cite{Tsori2004}.
\begin{figure}[!!!t]
\includegraphics[width=0.48\textwidth]{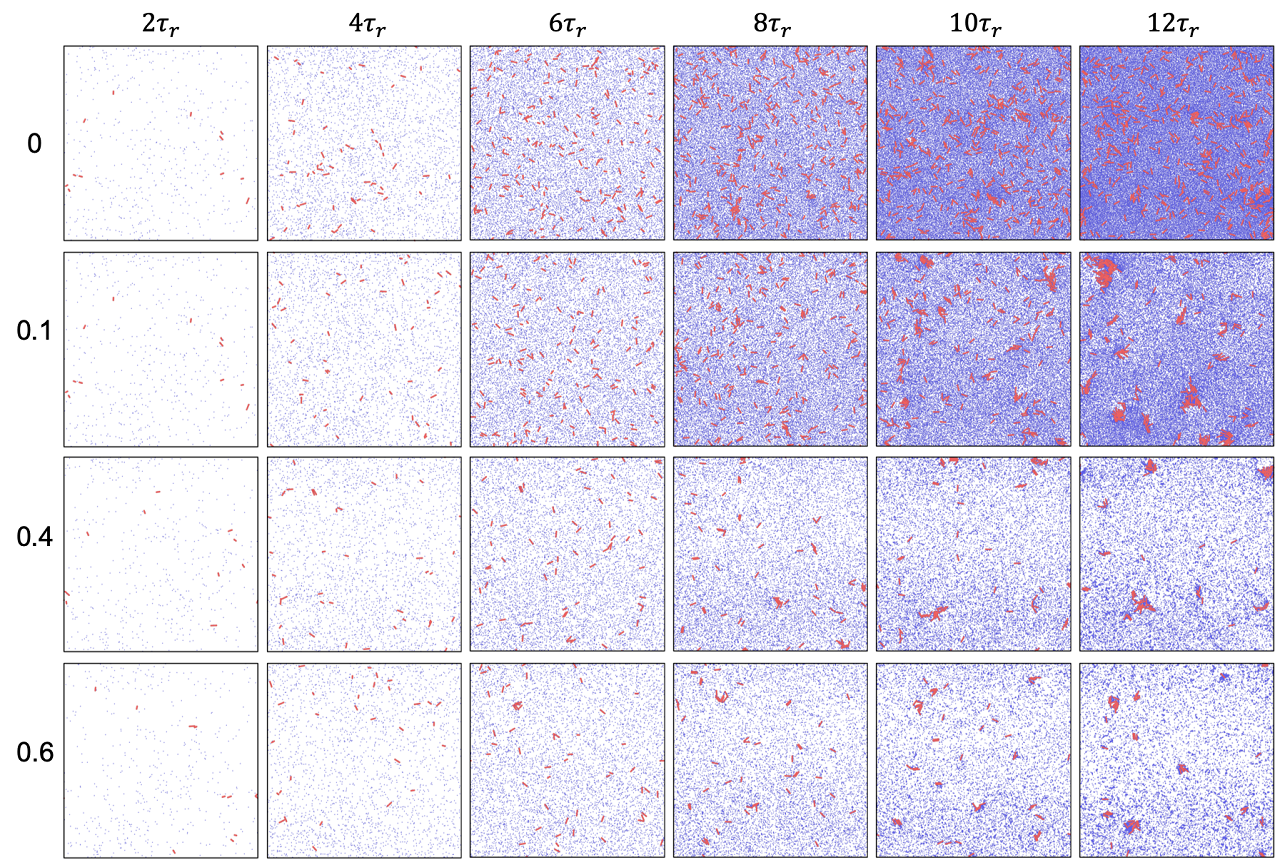}
\caption{Evolution of system of bacteria (red) which produce EPS (blue). 
The EPS particles can bond to each other, and to the bacteria, with probability $p_b$. 
Different rows correspond to different values of the bonding probability $p_b$, as indicated. }
\label{fig:overview_pb}
\end{figure}

To rationalize these results, we provide snapshots illustrating the time evolution of the investigated system in Fig.~\ref{fig:overview_pb}. 
In this figure, the columns correspond to different times, the rows to different values of the bond probability $p_b$, as indicated.
In all case, at long times, we do see the formation of small clusters of bacteria. 
These bacteria are glued together through the EPS particles. 
For small values of $p_b$, these clusters only form when there are many EPS particles in the systems.
Conversely, for a larger value of $p_b$, few EPS particles can glue the bacteria together.
Bacteria are therefore self-trapped by the EPS particles the produce~\cite{Tsori2004,Sengupta2009}.
The exponential dependence of the number of bacteria on $p_b$ observed in Fig.~\ref{fig:growth_pb}b is not simply recovered in a mean-field approximation, starting from rate equations from the total number of bacteria and the number of trapped bacteria. 
Spatial correlations, which are apparent in Fig.~\ref{fig:overview_pb}, appear therefore to play an important role in determining the size of the final population.

\section{From two- to three-dimensional microcolonies\label{sec:3d}}
\begin{figure}[!!t]
\centering
\includegraphics[width=0.5\textwidth]{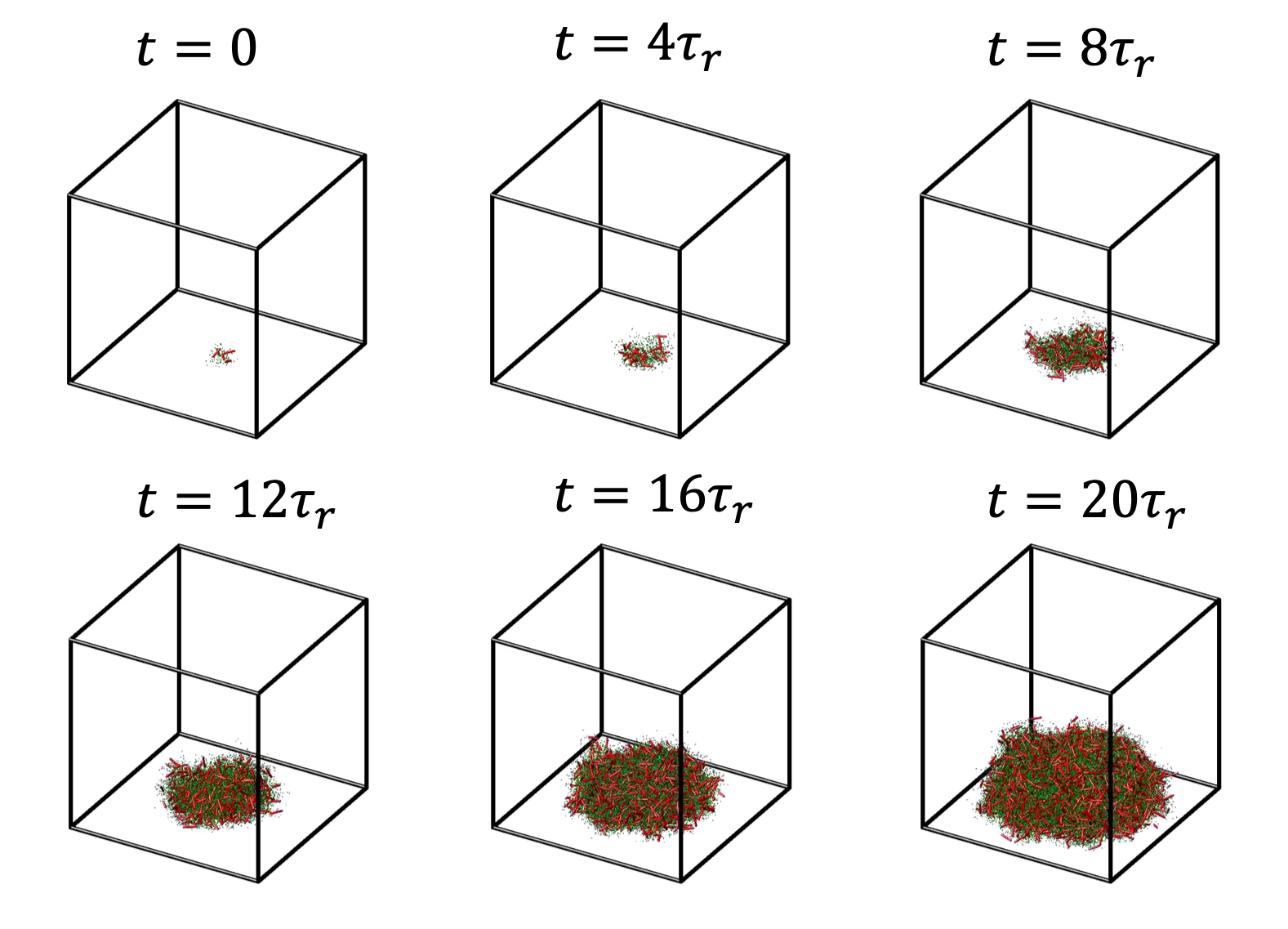}
\caption{Evolution of a three dimensional microcolony of not-motile bacteria. The microcolony develops with the bacteria embedded in an EPS gel matrix.}
\label{fig:3d}
\end{figure}
All investigations reported so-far have been restricted to the early stage formation of a biofilm, which is essentially a two-dimensional process. 
However, biofilms then develop as structured three dimensional aggregates. 
Here, without the aim of being quantitative, we demonstrate that the numerical approach we have developed is also able to describe this transition.
To this end, we extended the model to allow the bacteria to move in the vertical direction.

In the absence of EPS, the transition for two- to three-dimensional colonies has been suggested to originate from extrusion driven by the compression of the cells-\cite{Farrell2017,Grant2014} - alike in epithelial cell tissues.
In the presence of EPS, a different mechanism appear to be at work.
Indeed, while the bacteria are still on the plane, EPS particles move also in the vertical direction, and their polymerization leads to a three dimensional network. 
The stress induced in this network by the continuous growth and reproduction of the bacteria, leads to upward-forces acting on the bacteria, which force them out of the horizontal plane.
A small tilt of the bacterium is enough to seed the transition from a two to a three dimensional biofilm.

Fig.~\ref{fig:3d} illustrates the developing of a three dimensional biofilm, for non-motile bacteria.
Clearly, the bacteria result embedded in a growing EPS matrix. 
We leave to future studies the quantitative investigation of  three dimensional investigation, also because of their high computational cost.

\section{Conclusions}
In this manuscript, we have illustrated a computational model for the simulation of the early-stage biofilm formation.
The model reproduces results reported in previous numerical studies, such as the emergence of local nematic order, as well as the role of Psl trails.
Our model, however, shows for the first time that it is possible to describe in numerical setting the production of EPS as the growth of the extracellular matrix, in a coarse-grained fashion.

The main limitation of our model, and of related ones, appears the presence of many parameters. 
Specifically, the issue concerns the absence of a proper experimental measure of them, for most species. 
This renders a quantitative comparison with experimental results difficult. 
Nevertheless, the universality of the discussed phenomenology suggests that our model could suffice to pinpoint the key physical processes at work in the early-stage formation of a biofilm.

In this respect, our work suggests that not only the production of Psl trail~\cite{Zhao2013}, but also that of EPS, might induce the formation of microcolonies. 
Specifically, EPS leads to the formation of an extracellular matrix which traps the bacteria in what are de-facto microcolonies (see red regions in Fig.~\ref{fig:overview_pb}. 
Besides, we have originally observed that the incipient EPS matrix appears to foster the transition from a two- to a three-dimensional morphology.

\section*{Acknowledgements}
N.P. and M.P.C. acknowledge support from the Singapore Ministry of Education through the Academic Research Fund MOE2017-T2-1-066 (S), and are grateful to the National Supercomputing Centre (NSCC) for providing computational resources.


%
\end{document}